\documentclass[twocolumn,pre,superscriptaddress]{revtex4}

\usepackage{dcolumn}
\usepackage{amsmath}

\usepackage{graphicx}

\begin{document}

\renewcommand{\d}{{\rm d}}
\newcommand{\defn}{\textit}
\newcommand{\eref}[1]{(\ref{#1})}
\newcommand{\etal}{{\it{}et~al.}}
\newcommand{\br}{\mathbf{r}}

\newlength{\figurewidth}
\newlength{\pagewidth}
\setlength{\figurewidth}{0.95\columnwidth}
\setlength{\pagewidth}{1.9\columnwidth}
\setlength{\parskip}{0pt}
\setlength{\tabcolsep}{3.5pt}

\title{Optimal design of spatial distribution networks}
\author{Michael T. Gastner}
\affiliation{Santa Fe Institute, 1399 Hyde Park Road, Santa Fe, NM 87501}
\affiliation{Department of Physics, University of Michigan, Ann Arbor, MI 
48109}
\author{M. E. J. Newman}
\affiliation{Department of Physics, University of Michigan, Ann Arbor, MI 
48109}
\affiliation{Center for the Study of Complex Systems, University of
Michigan, Ann Arbor, MI 48109}

\begin{abstract}
We consider the problem of constructing public facilities, such as
hospitals, airports, or malls, in a country with a non-uniform population
density, such that the average distance from a person's home to the nearest
facility is minimized.  Approximate analytic arguments suggest that the
optimal distribution of facilities should have a density that increases
with population density, but does so slower than linearly, as the
two-thirds power.  This result is confirmed numerically for the particular
case of the United States with recent population data using two independent
methods, one a straightforward regression analysis, the other 
based on density dependent map projections.  We also consider
strategies for linking the facilities to form a spatial network, such as a
network of flights between airports, so that the combined cost of
maintenance of and travel on the network is minimized.  We show specific
examples of such optimal networks for the case of the United States.
\end{abstract}
\maketitle

\section{Introduction}

Suppose we are given the population density~$\rho(\br)$ of a country or
province, by which we mean the number of people per unit area as a function
of geographical position~$\br$.  And suppose we are charged with choosing
the sites of $p$ facilities, such as hospitals, post offices, supermarkets,
gas stations, or schools, so that the mean distance to the nearest facility
averaged over the population is minimized.  In most countries population
density is highly non-uniform, in which case a uniform distribution of
facilities would be a poor choice: it gains us little to build a lot of
facilities in sparsely populated areas.  A more sensible choice would be to
distribute facilities in proportion to population density, so that a region
with twice as many people has twice as many facilities.  But this
distribution too turns out to be suboptimal, because we also gain little by
having closely spaced facilities in the highly populated areas---when
facilities are closely spaced the typical person is not much further from
their second-closest facility than from their closest, so one or the other
can often be removed with little penalty and substantial savings.

As we will see, the ideal solution to this problem lies somewhere between
these two extremes, with the density of facilities increasing as the
two-thirds power of population density, a prediction that we verify using
simulations and visualizations based on cartograms, with actual population
data for the United States.  In addition, one is often interested in
connections between facilities, such as flights between
airports~\cite{Guimeraetal03} or transmission lines between power
stations~\cite{WattsStrogatz98}.  In the second half of this paper we
generate networks based on a simple model that optimizes network topology
with respect to the cost of maintaining and traveling across the network.
Depending on the benefit function chosen, we find structures ranging from
completely decentralized networks to hub-and-spoke networks.

\section{Optimal distribution of facilities}
\label{facdens}

We wish to distribute $p$ facilities over a two-dimensional area~$A$ such
that the objective function
\begin{equation}
f(\br_1\ldots\br_p) = 
\int_A\rho(\br)\min_{i\in\{1\ldots p\}}\!|\br-\br_i| \>\d^2r,
\label{optimfunc}
\end{equation}
is minimized.  Here $\{\br_1\ldots\br_p\}$ is the set of positions of
the facilities and $\rho(\br)$ is the population density within the
region~$A$ of interest.  This objective function is proportional to the mean
distance that a person will have to travel to reach their nearest facility.

Seemingly simple, this so-called \defn{$p$-median problem} has been shown
to be NP-hard~\cite{MegiddoSupowit84}, so in practice most studies rely
either on approximate numerical optimization or approximate analytic
treatments~\cite{Hochbaum97}.  A number of different approaches have been
used~\cite{MycielskiTrzeciakowski62,Viriakis69,Palmer73,Stephan77,Cameronetal02};
the calculation given here is essentially that of
Gusein-Zade~\cite{Gusein82}.

Our $p$ facilities naturally partition the area~$A$ into Voronoi cells.
(The Voronoi cell~$V_i$ for the $i$th facility is defined as the set of
points that are closer to $\br_i$ than to any other facility.)  Let us
define $s(\br)$ to be the area of the Voronoi cell to which the point $\br$
belongs.  In two dimensions a person living at point~$\br$ will on average
be a distance $g[s(\br)]^{1/2}$ from the nearest facility, where $g$ is a
geometric factor of order~1, whose exact value depends on the shape of the
Voronoi cell, but which will in any case drop out of the final result.  The
distance to the nearest facility averaged over all members of the
population is proportional to
\begin{equation}
f = g\int_A \rho(\br) [s(\br)]^{1/2}\>\d^2r,
\end{equation}
where we are making an approximation by neglecting variation of the
geometric factor~$g$ between cells.

The value of $s(\br)$ is constrained by the requirement that there be $p$
facilities in total.  Noting that $s(\br)$ is constant and equal to
$s(\br_i)$ within Voronoi cell~$V_i$, we see that the integral of
$[s(\br)]^{-1}$ over $V_i$ is
\begin{equation}
\int_{V_i} [s(\br)]^{-1} \>\d^2r = [s(\br_i)]^{-1} \int_{V_i} \d^2r = 1.
\end{equation}
Summing over all~$V_i$, we can then express the constraint on the number of
facilities in the form
\begin{equation}
\int_A [s(\br)]^{-1} \>\d^2r = p.
\label{constraint}
\end{equation}

Subject to this constraint, optimization of the mean distance~$f$ above
gives
\begin{equation}
{\delta\over\delta s(\br)} \biggl[ g\!\int_A \rho(\br) [s(\br)]^{1/2}\>\d^2r
  - \alpha \biggl( p - \int_A [s(\br)]^{-1} \>\d^2r \biggr) \biggr] = 0,
\end{equation}
where $\alpha$ is a Lagrange multiplier.  Performing the functional
derivatives and rearranging for $s(\br)$, we find $s(\br) =
[2\alpha/g\rho(\br)]^{2/3}$.  The Lagrange multiplier can be evaluated by
substituting into Eq.~\eref{constraint} and we arrive at the result
\begin{equation}
D(\br) = {1\over s(\br)}
       = p\,{[\rho(\br)]^{2/3}\over\int [\rho(\br)]^{2/3} \>\d^2r},
\label{optimum}
\end{equation}
where we have introduced the notation $D(\br)=[s(\br)]^{-1}$ for the
density of the facilities.

\begin{figure}
\begin{center}
\includegraphics[width=\columnwidth]{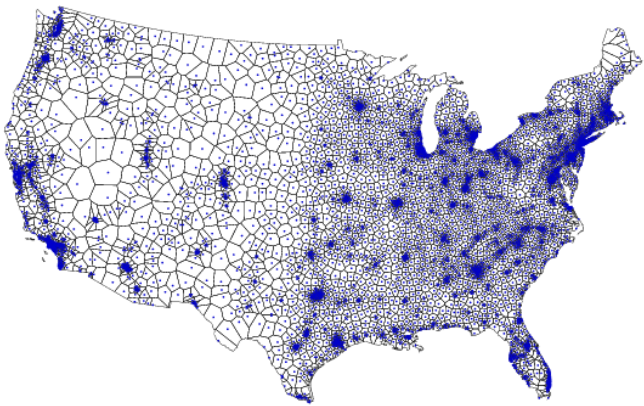}
\end{center}
\caption{Facility locations determined by simulated annealing and the
corresponding Voronoi tessellation for $p=5000$ facilities located in the
lower 48 United States, based on population data from the US Census for the
year 2000.}
\label{mapa}
\end{figure}

Thus, if facilities are distributed optimally for the given population
distribution, their density increases with population density but does so
slower than linearly, namely as a power law with
exponent~$\frac23$~\footnote{The analogous calculation can also be carried
out in general dimension~$d$; the corresponding exponent in that case is
$d/(d+1)$.}.  This density places most facilities in the densely populated
areas where most people live while still providing reasonable service to
those in sparsely populated areas where a strictly population-proportional
allocation might leave inhabitants with little or nothing.

The derivation above makes two approximations: it assumes that the
geometric factor $g$ is the same for all Voronoi cells and that $s(\br)$ is
a continuous function.  Neither assumption is strictly true, but we expect
them to be approximately valid if $\rho$ varies little over the typical
size of a Voronoi cell.  As a test of these assumptions, we have optimized
numerically the distribution of $p=5000$ facilities over the lower 48
states of the United States (Fig.~\ref{mapa}) using population data from
the most recent US Census~\cite{Census2003}, which counts the number of
residents within more than 8 million blocks across the study region.  To
create a continuous density function~$\rho$, we convolved these data with a
normalized Gaussian distribution of width 20~km.  The facility locations
were then determined by optimizing the full $p$-median objective
function~\eref{optimfunc} by simulated annealing~\cite{Gastner05}.

\begin{figure}
\begin{center}
\includegraphics[height=\columnwidth,angle=270]{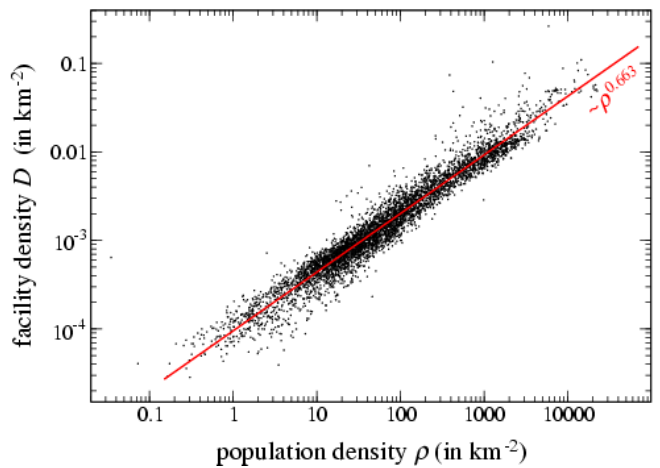}
\end{center}
\caption{Facility density $D$ from Fig.~\ref{mapa} versus population
density $\rho$ on a log-log plot.  A least-squares linear fit to the data
gives a slope of $0.663\pm0.002$ (solid line).}
\label{mapb}
\end{figure}

The relation $D\propto\rho^{2/3}$ can be tested as follows.  First, we
determine the Voronoi cell around each facility.  Then we calculate
$D(\br)$ as the inverse of the area of the corresponding cell and $\rho$ as
the number of people living in the cell divided by its area.
Figure~\ref{mapb} shows a scatter plot of the resulting data on
doubly-logarithmic scales.  If the anticipated $\frac23$-power relation
holds, we expect the data to fall along a line of slope~$\frac23$.  And
indeed a least-squares fit (solid line in the figure) yields a slope
$0.663(2)$, in good agreement with the theoretical prediction.

Some statistical concerns might be raised about this method.  First, we
used the Voronoi cell area to calculate both $D$ and~$\rho$, so the
measurements of $x$- and $y$-values in the plot are not independent, and
one might argue that a positive slope could thus be a result of artificial
correlations between the values rather than a real
result~\cite{Viningetal79}.  Second, it is known that estimating the
exponent of a power law such as Eq.~\eref{optimum} from a log-log plot can
introduce systematic biases~\cite{Newman05}.  In the next section, we
introduce an entirely different test of Eq.~\eref{optimum} that,
in addition to being of interest in its own right, suffers from neither of
these problems.

\section{Density-equalizing projections}
\label{cartogram}

\begin{figure*}
\begin{center}
\resizebox{\pagewidth}{!}{\includegraphics{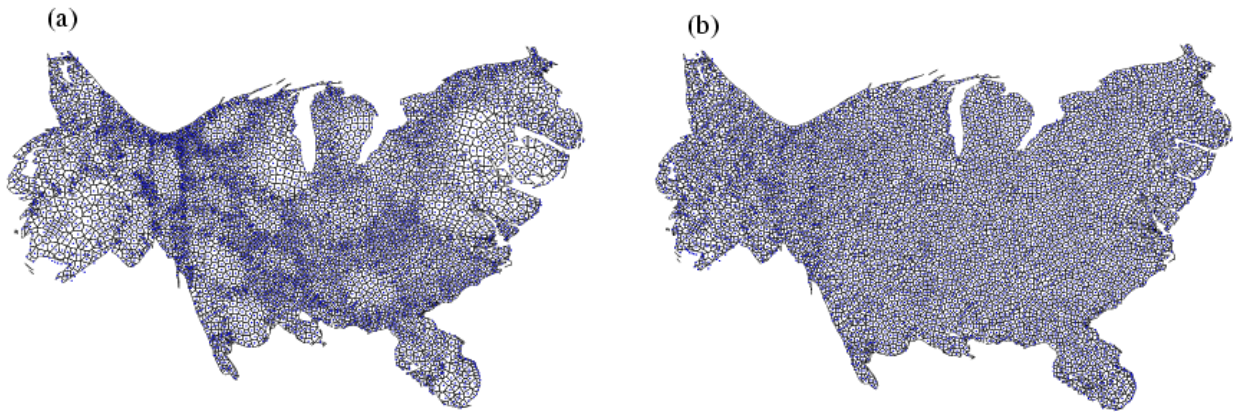}}
\end{center}
\caption{Near-optimal facility location on (a)~a cartogram equalizing the 
population density~$\rho$ and (b)~a cartogram equalizing~$\rho^{2/3}$.}
\label{cart}
\end{figure*}

If we neglect finite-size effects, it is straightforward to demonstrate
that optimally located facilities in a uniformly populated space lie on the
vertices of a regular triangular lattice~\cite{Loesch40}.  It has been
conjectured that for a non-uniform population there is a general class of
map projections that will transform the pattern of facilities to a
similarly regular structure~\cite{Bunge64}.  The obvious candidate
projections are population density equalizing maps or \defn{cartograms},
i.e.,~maps in which the sizes of geographic regions are proportional to the
populations of those
regions~\cite{GuseynTikunov94,Dorling96,Dent99,Tobler04}.  Densely
populated regions appear larger on a cartogram than on an equal-area map
such as Fig.~\ref{mapa}, and the opposite is true for sparsely populated
regions.  Since most facilities are located where the population density is
high, a cartogram projection will effectively reduce the facility density
in populated areas and increase it where the population density is low.
Therefore, one might expect that a cartogram leads to a more uniform
facility density than that shown in Fig.~\ref{mapa}.  And indeed some
authors have used population density equalizing projections as the basis
for facility location methods~\cite{Getis63,Rushton71}.

In Fig.~\ref{cart}a we show the facilities of Fig.~\ref{mapa} on a
population density equalizing cartogram created using the diffusion-based
technique of~\cite{GastnerNewman04}.  Although the \emph{population}
density is now equal everywhere, the \emph{facility} density is obviously
far from uniform.  A comparison between Fig.~\ref{mapa} and \ref{cart}a
reveals that we have overshot the mark since the facilities are now
concentrated in areas where there are few in actual space.

Equation~\eref{optimum} makes clear what is wrong with this approach.
Since $D$ grows slower than linearly with~$\rho$, a projection that
equalizes $\rho$ will necessarily overcorrect the density of facilities.
On the other hand, based on our earlier result, we would expect a
projection equalizing $\rho^{2/3}$ instead of $\rho$ to spread out the
facilities approximately uniformly.  Hence, one way to determine the actual
exponent for the density of facilities is to create cartograms that
equalize~$\rho^x$, $x\ge0$, and find the value of~$x$ that minimizes the
variation of the Voronoi cell sizes on the cartogram.  This approach does
not suffer from the shortcomings of our previous method based on the
doubly-logarithmic plot in Fig.~\ref{mapb}, since we neither use the
Voronoi cells to calculate the population density nor take logarithms.  One
might argue that the Voronoi cells on the cartogram are not equal to the
projections of the Voronoi cells in actual space, which is true---the cells
generally will not even remain polygons under the cartogram transformation.
The difference, however, is small if the density does not vary much between
neighboring facilities.

\begin{figure}
\begin{center}
\resizebox{\figurewidth}{!}{\includegraphics{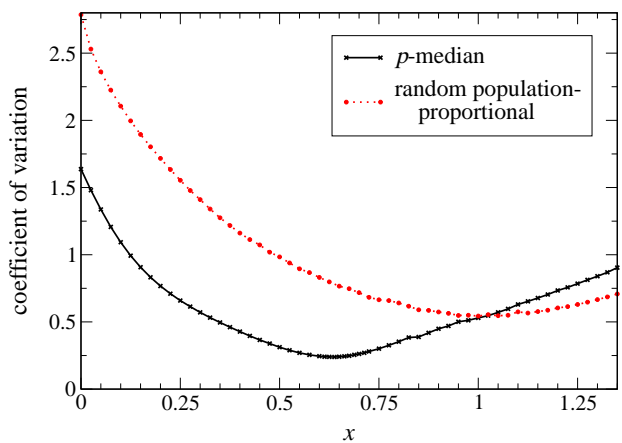}}
\end{center}
\caption{The coefficient of variation (i.e.,~the ratio of the standard
deviation to the mean) for Voronoi cell areas as they appear on a
cartogram, against the exponent $x$ of the underlying density $\rho^x$.}
\label{cellstd}
\end{figure}

In Fig.~\ref{cellstd} we show the measured coefficient of variation
(i.e.,~the ratio of the standard deviation to the mean) for Voronoi cell
sizes on $\rho^x$ cartograms as a function of the exponent~$x$ (solid
curve).  As the figure shows, the minimum is indeed attained at or close to
the predicted value of $x=\frac23$.  Figure~\ref{cart}b shows the
corresponding cartogram for this exponent.  This projection finds a
considerably better compromise between regions of high and low population
density than either Fig.~\ref{mapa} or~\ref{cart}a.

For comparison, we have also made the same measurement for $5000$ points
distributed randomly in proportion to population.  Since the density of
these points is by definition equal to~$\rho$, we expect the minimum
standard deviation of the cell areas to occur on a cartogram with $x=1$.
Our numerical results for this case (dotted curve in Fig.~\ref{cellstd})
agree well with this prediction.  Comparing the solid and the dotted curves
in the plot, we see that not only the positions of the minima differ, but
also the minimal values themselves.  The lower standard deviation for the
$p$-median distribution indicates that optimally located facilities are not
randomly distributed with a density $\propto\rho^{2/3}$.  Instead, the
optimally located facilities occupy space in a relatively regular fashion
reminiscent of the triangular lattice of the uniform population
case~\cite{Loesch40,Christaller80}.

\section{Optimal networks of facilities}
\label{netw}

In many cases of practical interest, finding the optimal location of
facilities is only half the problem.  Often facilities are interconnected
forming networks, such as airports connected by flights or warehouses
connected by truck deliveries.  In these cases, one would also like to find
the best way to connect the facilities so as to optimize the performance of
the system as a whole.

Consider then a situation in which our facilities form the nodes or
vertices of a network and connections between them form the edges.  The
efficiency of this network, as we will consider it here, depends on two
factors.  On the one hand, the smaller the sum of the lengths of all edges,
the cheaper the network is to construct and maintain.  On the other hand,
the shorter the distances through the network between vertices, the faster
the network can perform its intended function (e.g.,~transportation of
passengers between nodes or distribution of mail or cargo).  These two
objectives generally oppose each other: a network with few and short
connections will not provide many direct links between distant points and
paths through the network will tend to be circuitous, while a network with
a large number of direct links is usually expensive to build and operate.
The optimal solution lies somewhere between these extremes.

Let us define $l_{ij}$ to be the shortest geographic distance between two
vertices $i$ and $j$ measured along the edges in the network.  If there is
no path between $i$ and~$j$, we formally set $l_{ij}=\infty$.  Introducing
the adjacency matrix $\mathbf{A}$ with elements $A_{ij}=1$ if there is an
edge between $i$ and~$j$ and $A_{ij}=0$ otherwise, we can write the total
length of all edges as
\begin{equation}
T = \sum_{i<j} A_{ij}l_{ij}.
\label{defst}
\end{equation}
We assume this quantity to be proportional to the cost of maintaining the
network.  Clearly this assumption is only approximately correct; networked
systems in the real world will have many factors affecting their
maintenance costs that are not accounted for here.  It is however the
obvious first assumption to make and, as we will see, can provide us with
good insight about network structure.

The typical cost of shipping a commodity or traveling through the network
depends on the distances~$l_{ij}$ as well as the amount of traffic~$w_{ij}$
(e.g.,~weight of cargo, number of passengers, etc.)\ that flows between
vertices $i$ and~$j$.  In a spirit similar to our assumption about
maintenance costs, we assume that the total travel cost is given by
\begin{equation}
Z = \sum_{i<j} w_{ij}l_{ij}.
\label{Z}
\end{equation}
We assume that $w_{ij}$ is proportional to the product of populations in
the Voronoi cells $V_i$ and $V_j$ around $i$ and~$j$, so that
\begin{equation}
w_{ij} = \int_{V_i}\rho(\br)\>\d^2r\int_{V_j}\rho(\br')\>\d^2r'
\label{weight}
\end{equation}
in appropriate units.  And the total cost of running the network is
proportional to the sum $T+\gamma Z$ with $\gamma\ge0$ a constant that
measures the relative importance of the two terms.  Then the optimal
network is the one minimizing this
sum~\cite{BillheimerGray73,LosLardinois82}.

Using again the conterminous United States as an example, we have first
determined the optimal placement of $p=200$ facilities which we then try to
connect together optimally.  The number of edges in the network depends on
the parameter~$\gamma$.  If $\gamma\to 0$, the cost of travel~$Z$ vanishes
and the optimal network is the one that simply minimizes the total length
of edges.  That is, it is the minimum spanning tree, with exactly $p-1$
edges between the $p$ vertices.  Conversely, if $\gamma\to\infty$ then
$Z$~dominates the optimization, regardless of the cost~$T$ of maintaining
the network, so that the optimum is a fully connected network or clique
with all $\frac12 p(p-1)$ possible edges present.  For intermediate values
of~$\gamma$, finding the optimal network is a non-trivial combinatorial
optimization problem, for which we can derive good, though usually not
perfect, solutions using again the method of simulated
annealing~\cite{Gastner05}.

\begin{figure*}
\begin{center}
\resizebox{\pagewidth}{!}{\includegraphics{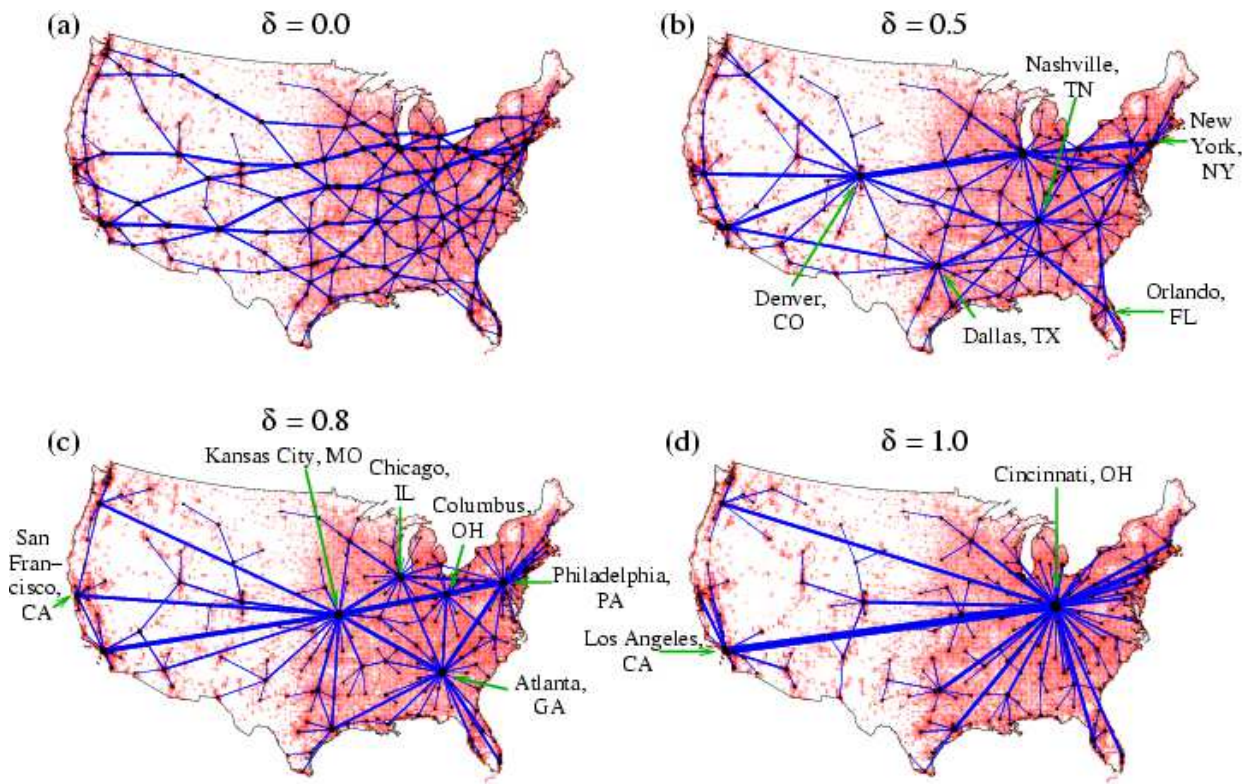}}
\end{center}
\caption{Optimal networks for the population distribution of the United
States with $p=200$ vertices for different values of $\delta$ and
with~$\gamma=10^{-14}$.}
\label{netwpmed}
\end{figure*}

There is, however, another complicating factor.  In Eq.~\eref{Z} we assumed
that travel costs are proportional to geometric distances between vertices,
which is a plausible starting point.  In a road network, for example, the
quickest and cheapest route is usually not very different from the shortest
route measured in kilometers.  But in other networks travel costs can also
depend on the number of legs in a journey.  In an airline network, for
instance, passengers often spend a lot of time waiting for connecting
flights, so that they care both about the total distance they travel and
the number of stopovers they have to make.  Similarly, the total time
required for an Internet packet to reach its destination depends on two
factors, the propagation delay proportional to the physical distance
between vertices (computers and routers) and the store and forward delays
introduced by the routers, which grow with the number of intermediate
vertices.

To account for such situations, we generalize our definition of the length
of an edge and assign to each edge an effective length
\begin{equation}
\tilde{l}_{ij} = (1-\delta) l_{ij}+\delta
\label{efflngth}
\end{equation}
with $0\leq\delta\leq 1$.  The parameter $\delta$ determines the user's
preference for measuring distance in terms of kilometers or legs.  Now we
define the effective distance between two (not necessarily adjacent)
vertices to be the sum of the effective lengths of all edges along a path
between them, minimized over all paths.  The travel cost is then
proportional to the sum of all effective path lengths
\begin{equation}
Z = \sum_{i<j} w_{ij}\tilde{l}_{ij},
\label{Zdelta}
\end{equation}
and the optimal network for given $\gamma$ and $\delta$ is again the one
that minimizes the total cost $T+\gamma Z$.  (Since the second term in
Eq.~\eref{efflngth} is dimensionless, we normalize the length appearing in
the first term by setting the average ``crow flies'' distance between a
vertex and its nearest neighbor equal to one.)

In Fig.~\ref{netwpmed} we show the results of the application of this
process to the lower 48 United States.  When $\delta=0$ passengers (or
cargo shippers) care only about total kilometers traveled and the optimal
network strongly resembles a network of roads, such as the US interstate
network.  As~$\delta$ increases the number of legs in a journey starts
playing a more important role and the approximate symmetry between the
vertices is broken as the network begins to form hubs.  Around $\delta=0.5$
we see networks emerging that constitute a compromise between the
convenience of direct local connections and the efficiency of hubs, while
by $\delta=0.8$ the network is dominated by a few large hubs in
Philadelphia, Columbus, Chicago, Kansas City, and Atlanta that handle the
bulk of the traffic.  On the highly populated Californian coast, two
smaller hubs around San Francisco and Los Angeles are visible.  In the
extreme case $\delta=1$, where the user cares only about number of legs and
not about distance at all, the network is dominated by a single central hub
in Cincinnati, with a few smaller local hubs in other locations such as Los
Angeles.

\section{Conclusions}
\label{concl}

We have in this paper studied the problem of optimal facility location,
also called the $p$-median problem, which consists of choosing positions
for $p$ facilities in geographic space such that the mean distance between
a member of the population and the nearest facility is minimized.  Analytic
arguments indicate that the optimal density of facilities should be
proportional to the population density to the two-thirds power.  We have
confirmed this relation by solving the $p$-median problem numerically and
projecting the facility locations on density-equalizing maps.  We have also
considered the design of optimal networks to connect our facilities
together.  Given optimally located facilities, we have searched numerically
for the network configuration that minimizes the sum of maintenance and
travel costs.  A simple two-parameter model allows us to take different
user preferences into account.  The model gives us intuition about a number
of situations of practical interest, such as the design of transportation
networks, parcel delivery services, and the Internet backbone.

\begin{acknowledgments}
The authors thank the staff of the University of Michigan's Numeric and
Spatial Data Services for their help.  This work was funded in part by the
National Science Foundation under grant number DMS--0234188 and by the
James S. McDonnell Foundation.
\end{acknowledgments}

\end{document}